\documentstyle[12pt]{article}
\newcommand{\Jpsi}{J\!/\!\psi} 
\begin{document}
\begin{center} 
{{\bf POLARIZATION EFFECTS IN \\
DRELL-YAN TYPE PROCESSES 
$h_1+h_2\rightarrow (W,Z,\gamma^{\ast},J\!/\!\psi)+X$}}\\[2mm]

{E.~MIRKES\footnote{Talk
  presented by E. Mirkes at the DPF94 Meeting, Albuquerque,
New Mexico, USA; August 2-6, 1994.}\\
{\em Physics Department, University of Wisconsin, Madison,
WI 53706, USA }\\
\vspace{0.3cm}
and\\
\vspace*{0.3cm}
J.~OHNEMUS\\
{\em Physics Department, University of California, Davis, CA 95616, USA
     }}
\end{center}

\setlength{\baselineskip}{2.6ex}
 
\begin{center}
\parbox{13.0cm}
{\begin{center} ABSTRACT \end{center}
{\small \hspace*{0.3cm}
The measurement of the angular distribution of leptons  from the decay
of a $W$, $Z$, $\gamma^{\ast}$, or $\Jpsi$ produced at high
transverse momentum in hadronic collisions
provides a detailed test of the production and decay  mechanism of the 
spin one state.  In the absence of cuts on the final state leptons, the  lepton
angular distribution in the lepton pair rest frame is determined by
the polarization of the 
spin one state. 
At leading order in perturbative QCD the general structure
of the decay lepton distribution arising from a 
$W,Z\,\,  [\gamma^{\ast},\Jpsi]$
is controlled by six [four] invariant structure functions.
In the presence of cuts, the lepton angular
distributions are dominated by kinematic effects rather than
polarization effects.
We present Monte Carlo studies for Tevatron energies 
and discuss  how polarization effects can be highlighted in the presence
of cuts.}}
\end{center}
 
The physics of a gauge boson or a $\Jpsi$ produced at high transverse
momentum at hadron colliders is a rich source of information on many
aspects of the physics of the standard model \cite{cs,npb,jim,jpsi}.
For gauge bosons or $\Jpsi$'s produced with transverse momentum $p_T$,
the event plane spanned by the beam and
the spin-one state  momentum directions provides a convenient reference 
plane for studying the angular distributions of the decay leptons.
In leading order QCD 
the  angular distribution of the leptons from a gauge boson 
$V$ [$V=W,Z,\gamma^{\ast}$]
has the general form:
\begin{eqnarray}
\frac {d\sigma}{d p_{T}^{2}\,dy\, d\cos\theta \,d\phi} 
&=&  \frac{3}{16\pi}\,
\frac{d\sigma^{V}_{U+L}}{ d p_{T}^{2}\,dy}\,\,
           \,\left[  (1+\cos^{2}\theta)\,\, \nonumber 
               +\,\, \frac{1}{2}A_{0}^V \,\,\, (1-3\cos^{2}\theta) 
 \right.\\[2mm]
&&   \left.
\hspace{-3.8cm} +  \,\,   A_{1}^V  \,\,\,\sin 2\theta \cos\phi \,\,
\,\, + \,\,   \frac{1}{2}A_{2}^V  \,\,\,\sin^{2}\theta\cos 2\phi\nonumber  
\,\,+  \,\,   A_{3}^V  \,\,\,\sin \theta \cos\phi \,\,
\,\, + \,\,   A_{4}^V  \,\,\,\cos\theta  
\,\,  \right] \>,
  \label{ang}    
  \end{eqnarray}
where $\theta$ and $\phi$ denote the polar and azimuthal angle of the
decay leptons in the lepton pair rest frame.
The coefficients $A_i$ are functions of the transverse momentum $p_T$
and the rapidity $y$ of the gauge boson $V$.
For $V=\gamma^{\ast}$, the parity violating coefficients $A_3$ and
$A_4$ are zero.
The unpolarized differential production cross section is denoted by
${\sigma}_{U+L}$ whereas the coefficients $A_i$ characterize the 
polarization of the spin-one state, {\it e.g.}, the cross section contribution of
the longitudinal  polarization is denoted by $A_0$, the 
transverse-longitudinal interference contribution by $A_1$, 
the transverse interference cross section by $A_2$, {\it etc}
[all with respect to the $z$-axis of the chosen lepton pair rest 
frame]. \cite{npb,jim}
With the replacements $V\rightarrow \Jpsi$ and $A_{3,4}\rightarrow 0$, 
the general form of 
Eq.~(\ref{ang}) is also valid for
the decay lepton distribution arising from the decay
of  a $\Jpsi$ produced at high transverse momentum in hadronic 
collisions \cite{jpsi}.

The coefficients $A_i$ are dependent on
the choice of the $z$-axis in the lepton pair rest frame.
We will present results here for the Collins-Soper 
\cite{cs} (CS) frame. In this frame the $z$-axis bisects the angle between
the proton and negative anti-proton direction.
 For $W$ decay, this frame
has the unique advantage that the polar and azimuthal angles $\theta$
and $\phi$ can be reconstructed   modulo a sign ambiguity in
$\cos\theta$ without information on the longitudinal momentum of the
neutrino  \cite{npb,jim}. 
The complete
next-to-leading-order (NLO) corrections to the parity conserving
coefficients were calculated in Ref.~2 and  were found to be
fairly small [less than 10\%].
It is thus  sufficient to use LO matrix elements in the Monte Carlo 
study presented in this paper.
Figure~1 shows the normalized  $\phi$ and $\cos\theta$
distributions for the leptons from the decay of a $Z$ boson for
four bins in the transverse momentum of the $Z$ boson. 
No acceptance cuts
have been applied to the leptons.
For a more realistic analyses, one has to impose acceptance
cuts and energy resolution smearing to the leptons.
When imposing the cuts $p_T(l)> 25$ GeV and $|y(l)|<1$ to the leptons,
the shapes of the lepton angular
distributions are dominated by kinematic effects and the residual
dynamical effects from the gauge boson polarization are small. 
 Polarization effects can be maximized
by minimizing the cuts, however, this strategy is severly limited
since cuts are needed to reject  background events.
Therefore, to regain sensitivity to the
polarization effects in the presence of large kinematic cuts, we
propose to divide the experimental distributions by the Monte Carlo
distributions obtained using isotropic gauge boson decay.
The large effects of the cuts [and smearing in the $W$ case \cite{jim}]
are expected to almost cancel in this ratio.
Figure~2 shows the
ratio of  the $\phi$ and $\cos\theta$
distributions  for the $Z$ with full polarization to the corresponding
distributions obtained with isotropic $Z$ decay for the same bins as
in Fig.~1.
Energy resolution smearing and the cuts $p_T(l)>25$ GeV and $|y(l)|<1$ 
are included.
The ratios contain most of the polarization dependence seen in Fig.~1.
A  detailed Monte Carlo analyses of the decay angular distributions 
of  $W$'s and $Z$'s produced at Tevatron energies is
presented in Ref.~3.
Analytical and numerical results [calculated in the
nonrelativistic bound state model] for the decay lepton distribution
of high $p_T$ $\Jpsi$'s can be found in 
Ref.~4.
All  distributions are fairly insensitive to the scales in the
strong coupling constant  and the
choice of the parton distribution functions.

\bibliographystyle{unsrt}

\begin{center}
{\bf FIGURE CAPTIONS}
\end{center}
\begin{itemize}
\item[{\bf Fig. 1}]
a) Normalized $\phi$ and b) normalized $\cos\theta$ distributions  of
the negatively charged lepton from $Z$ boson decay in the CS frame.
Results are shown for four bins in $p_T^{}(Z)$:\\
10~GeV $<\, p_T^{}(Z)\, < 20$~GeV (solid),\\
20~GeV $<\, p_T^{}(Z)\, < 30$~GeV (dashed),\\
30~GeV $<\, p_T^{}(Z)\, < 70$~GeV (dots),\\
70~GeV $<\, p_T^{}(Z)\, $ (dot-dashed).\\
No cuts or smearing have been applied.
\item[{\bf Fig. 2}]
Ratios of distributions obtained with full polarization effects to
those obtained with isotropic decay of the $Z$ boson. Parts a) and b)
are the ratios for the $\phi$ and $\cos\theta$ distributions,
respectively. Energy resolution smearing and the cuts $p_T^{}(l) >
25$~GeV and $|y(l)| < 1$ are included.
\end{itemize}

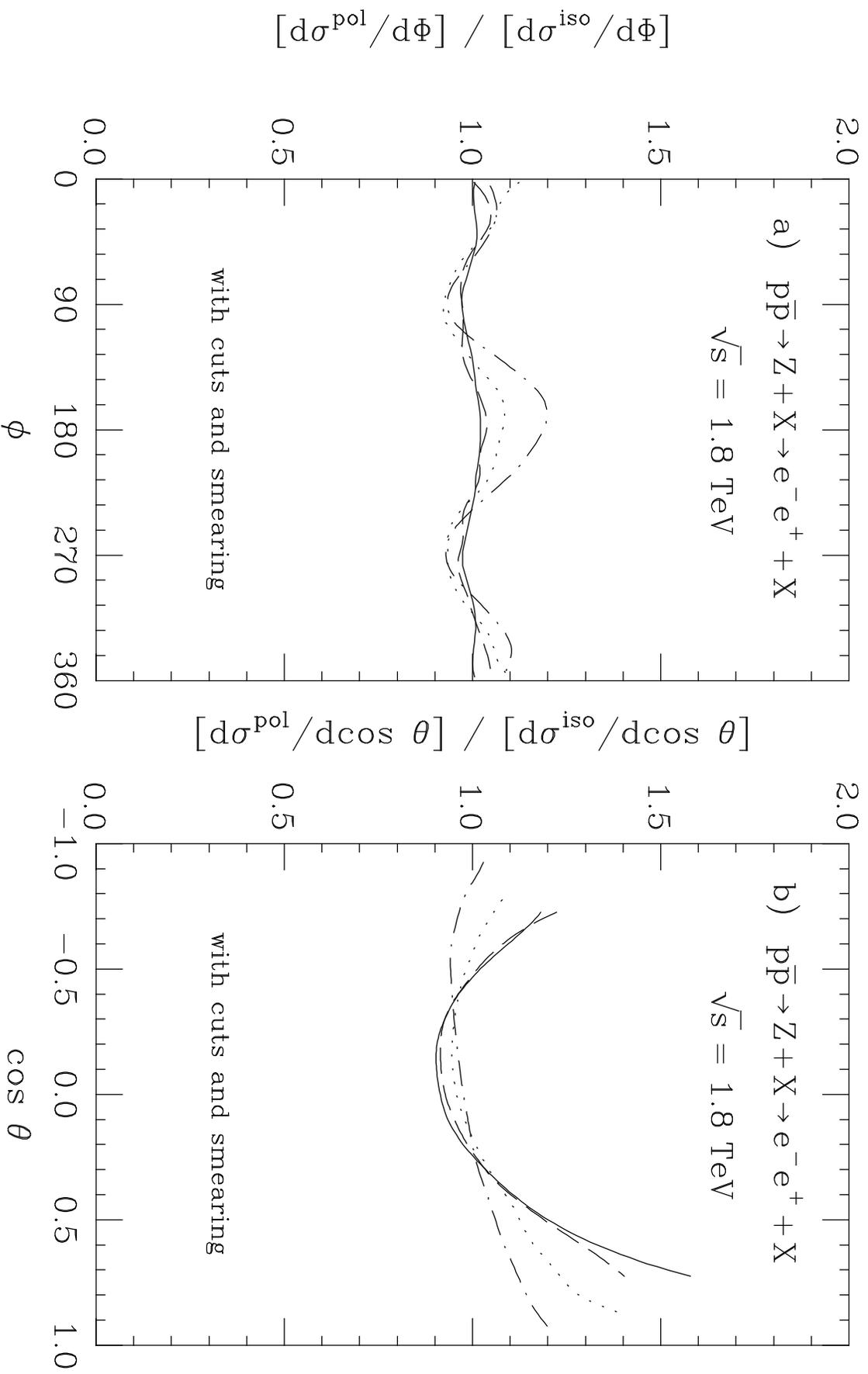
\end{document}